# HTS Dipole Magnet with a Mechanical Energy Transfer in the Magnetic Field

Vladimir Kashikhin, Daniele Turrioni

*Abstract*— **There were designed and successfully tested at Fermilab several high temperature superconducting (HTS) model magnets for particle accelerators. Some of them worked in a persistent current mode continuously generating magnetic field in the iron dominated magnet gap. In the paper was investigated a novel HTS dipole magnet concept with a mechanical energy transfer in the magnetic field. To pump the energy in the superconducting HTS dipole magnet used a detachable magnetizer. The HTS dipole magnet was build and successfully tested at a liquid nitrogen temperature. Discussed the magnet design, test results, the proposed approach limits, and efficiency.**

## I. Introduction

ADVANCES in the high temperature superconductors (HTS) attracted an interest of accelerator magnet designers. Most of researchers concentrated on the high magnetic field applications because HTS capable to work at much higher magnetic fields than widely used NbTi or Nb3Sn superconductors. Besides HTS magnet systems is more difficult to protect from quenches and high current HTS cables not readily available from the industry. Nevertheless, at that time could be used advantages of HTS magnet technology for relatively low field accelerator iron dominated magnets working at elevated temperatures up to 77 K. At Fermilab were built and tested several HTS magnet models [1] – [3]. Some of them worked in a persistent current mode. In this paper described the HTS dipole magnet working in a persistent current mode. To energize this magnet used a removable magnetizer which transfer the energy in the short-circuited HTS coil. The magnet was successfully tested in a liquid nitrogen. Test results presented and discussed.

## II. Short-Circuited HTS Coils

Most known high temperature superconducting magnet coils wound from tape type superconductors or HTS cables. Because a quench propagation in HTS is very slow, it is difficult to protect them from a local overheating. Besides it is also difficult to detect a quench because for the short length of HTS transferred in the normal condition the voltage drop in this area is very low. In [4], [5] was proposed a way to make short-circuited HTS multi turn coils from the tape type superconductor having slit in the middle besides both ends. In [2] –[3] this approach was updated by using an assembly of short-circuited HTS loops working in parallel, as shown in Fig. 1. In this case, the coil is self-protected against quenches because of current sharing induced by a mutual field and resistive coupling.

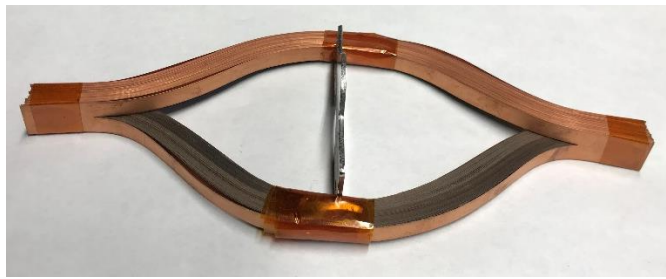

Fig. 1. HTS coil assembled from parallel short-circuited loops.

## III. HTS Dipole Magnetic Design

The dipole magnet design based on the mechanical energy transfer concept described in [6]. The investigated HTS dipole magnet model has C-type configuration and consists of an iron core with HTS short-circuited coils (See Fig. 1) mounted on a core back leg as shown in Fig. 2. To pump the magnetic field energy in the magnet used detachable magnetizer having a conventional copper coil. HTS dipole magnet with attached magnetizer shown in Fig. 3. The magnetic field was simulated using OPERA3D software [7].

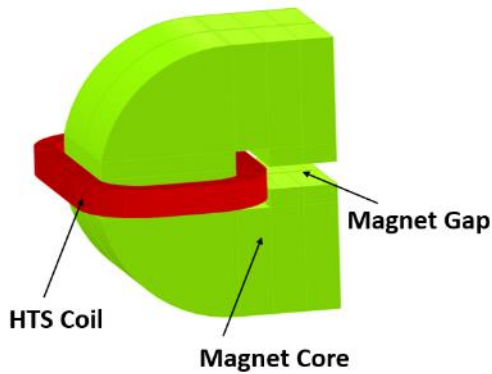

Fig. 2. HTS Dipole magnet model.

The Magnetizer initially attached to the dipole magnet core to excite the magnetic flux in the closed magnet core. For that needed relatively low current $I1$ in the copper coil.

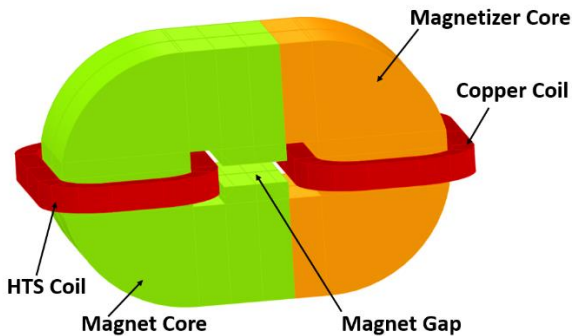

Fig. 3. HTS Dipole magnet (left) with the Magnetizer (right).

To induce the current in HTS coil the Magnetizer should be mechanically removed from the magnet. Fig. 4 shows how

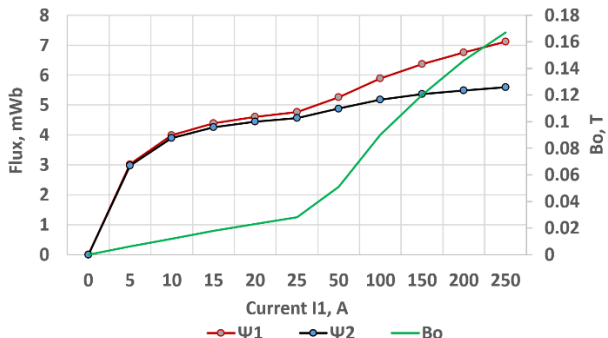

magnetic fluxes depends on the primary copper coil currents.

Fig. 4. HTS Dipole magnet with the attached Magnetizer. The copper coil flux $\Psi1$, the HTS coil flux $\Psi2$, and the gap field $Bo$.

One can see that the iron core saturation started after 10 A current and after 25 A more flux goes through the gap sharply increasing the magnet gap field. Fig. 5 shows magnetic fluxes generated by primary and secondary coils. In an agreement with Lentz law the induced in HTS coil current must cancel the flux generated by the primary coil.

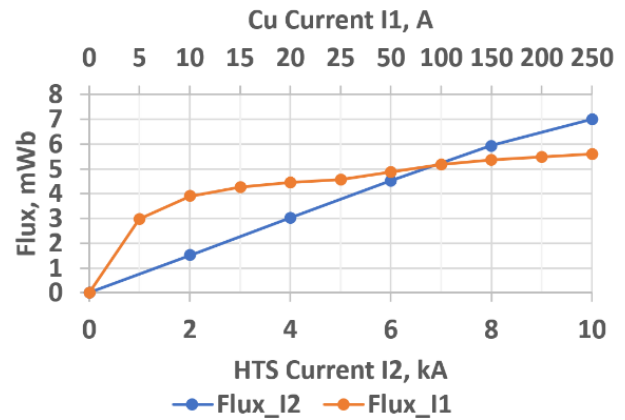

Fig. 5. The magnetic flux for HTS coil and Cu coil for the closed magnet core.

So, for example, the flux 3 mWb could be generated by 5 A primary coil current, which induces 4 kA current in HTS coil. This 4 kA current produces 0.5 T field in the magnet gap (See Fig. 6).

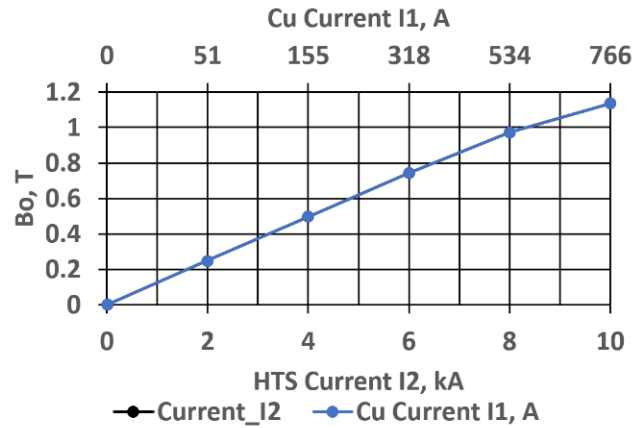

Fig. 6. The magnet gap field for initial primary $I1$ current and induced HTS currents $I2$.

IV. HTS MAGNET DESIGN AND FABRICATION

The HTS dipole magnet has a conventional iron-dominated accelerator magnet configuration having a gap of 10 mm. The magnet core geometry shown in Fig. 7. The core assembled from 6.35 mm thick low carbon steel plates bolted in a longitudinal direction. Two HTS coils assembled from HTS loops as shown in Fig. 1 and mounted around the core back leg (See Fig. 8). Inside the magnet gap were mounted two Hall probes to monitor the gap field during operations. To the HTS coil end was attached a CERNOX temperature sensor.

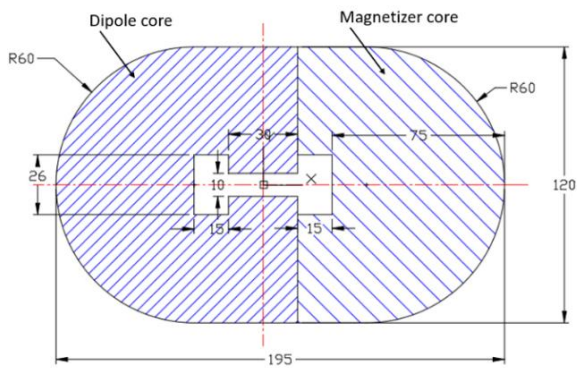

Fig. 7. Magnet core cross-section.

The dipole magnet main parameters are shown in Table 1, and the magnet view in Fig. 8.

TABLE I
DIPOLE MAGNET DESIGN PARAMETERS

| Parameter | Unit | Primary copper coil | Secondary HTS coil |
|---|---|---|---|
| Dipole magnet gap | mm | 10 | |
| Coil number of turns/loops | | 20 | 112 |
| Conductor | | Copper | HTS |
| Conductor dimensions | mm | 2x2 | 0.1x12* |
| Peak coil total current | A | 3800 | 6000 |
| Peak field in the gap | T | 0.74 | |
| Magnet length | mm | 64 | |
| Outer yoke dimensions | mm | 120 x 195 | |

*12 mm superconductor from SuperPower split to form 6 mm wide loops.

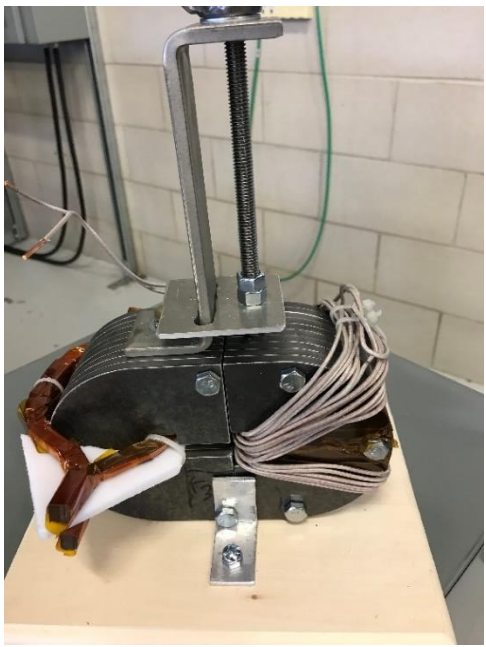

Fig. 8. Assembled HTS dipole magnet with the Magnetizer.

## V. HTS DIPOLE MAGNET TEST

The HTS Dipole magnet was tested in the LN$_2$ bath. Fig 9 shows the magnet which had just been removed from the LN$_2$ bath. HTS coils currents continue to circulate during several minutes of warming up process. With the coil temperature rises the conductor resistance and currents slowly decays. The heat evenly distributed in HTS loops which protects the coil from the heat concentration in small areas which usually happened in the multi turn coils.

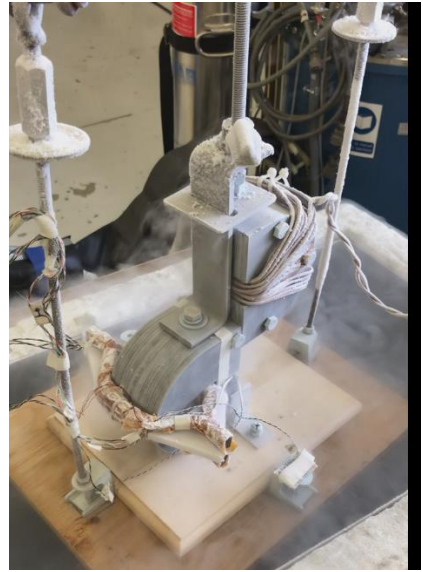

Fig. 9. Magnet just removed from the LN$_2$ bath, but HTS coil currents continue to circulate.

Fig. 10 the dipole magnet operation started from the initial 25 A in the Magnetizer, cooling down to 77 K, and pumping the mechanical energy in the magnetic field by lifting the Magnetizer. There were 6 steps with 20 mm (10 handle rotations) travel up with short ~ 1 min stops.

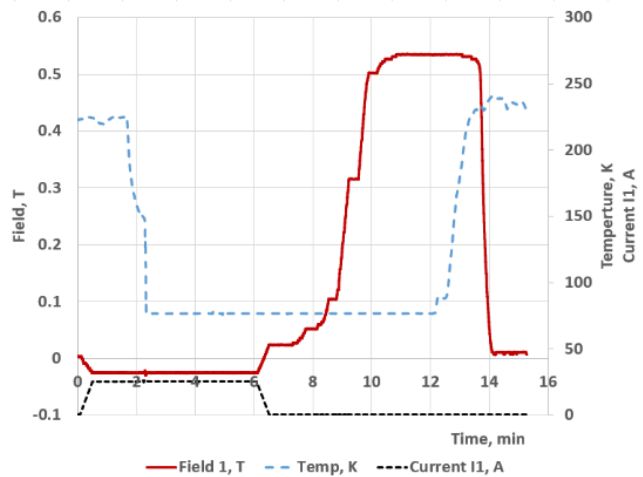

Fig. 10. The whole cycle of dipole magnet operation. Peak field in the gap 0.54 T.

This test was repeated for 50 A, 100 A, and 190 A initial Magnetizer currents. Fig. 11 shows the magnetic field variation for these currents. One could see that the large field increase is in the range of 40 mm – 80 mm of Magnetizer displacement when opens the dipole magnet gap.

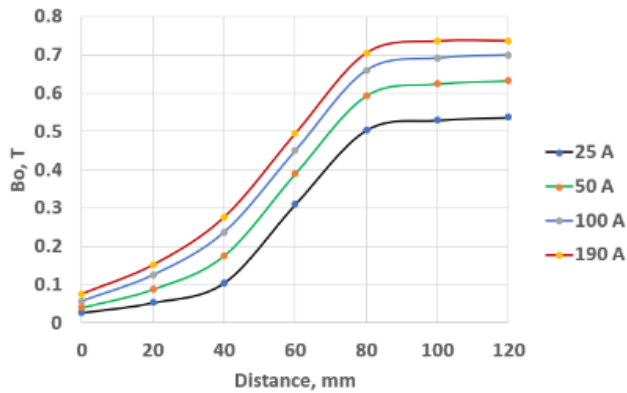

Fig. 11. Flux density variation in the magnet gap for different vertical Magnetizer displacements and initial copper coil currents $I1$.

The peak field 0.75 T reached at the initial $I1$ current of 190 A as shown in Fig. 12 and limited by the iron core saturation.

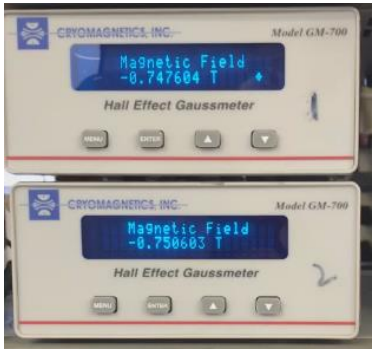

Fig. 12. Two Hall probes used to monitor the magnetic field in the magnet gap. They showed the 0.75 T peak field reached during tests.

Fig, 13 shows comparison of the primary coil ampere-turns $Iw1$ and induced HTS total current $Iw2$.

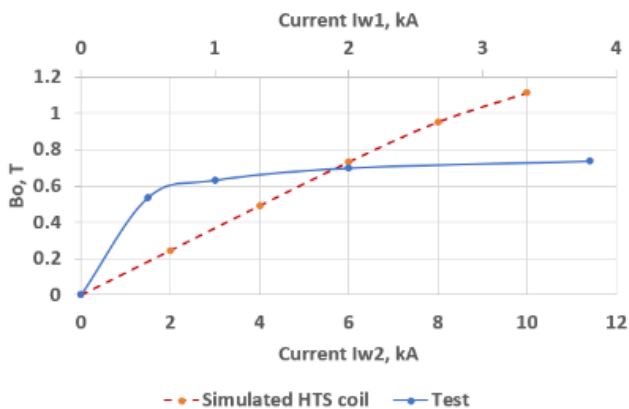

Fig. 13. Flux density variation in the magnet gap for the simulated HTS coil current $Iw2$ and the initial copper coil total current $Iw1$ in the Magnetizer.

The long-term test was performed o confirm that this type of magnet capable work many hours without current decay as shown in Fig. 14.

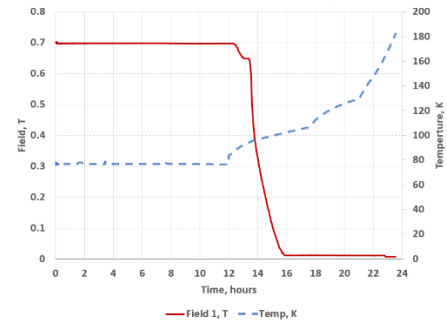

Fig. 14. Long-term test with HTS current and zero current $I1$ and with the lifted Magnetizer.

The HTS coils current circulated 12 hours generating 0.7 T field without decay. The initial Magnetizer coil current was 100 A. After 12 hours these coils slowly warmed up during ~2 hours as the $LN_2$ in the bath was evaporated and currents decayed to zero.

## Conclusion

The HTS dipole magnet was designed, fabricated, and successfully tested. There was confirmed an approach of using mechanical energy transfer in the magnetic field [6] with following magnet work with a persistent current in HTS coils and disconnected power source. During the test was confirmed that HTS coils of this type are self-protected. This type of magnets could generate permanently stable magnetic field with removed Magnetizer.

## Acknowledgment

The authors would like to thank the FNAL APS-TD team and management for the support of this work.